\documentclass[twocolumn,showpacs,preprintnumbers]{revtex4}
\usepackage{amssymb}

\usepackage{graphicx}
\usepackage{dcolumn}
\usepackage{bm}

\begin{document}

\title{Amplification of Local Instabilities in a Bose-Einstein Condensate with Attractive Interactions}

\begin{abstract}
We study the collapse of large homogeneous Bose-Einstein
condensates due to intrinsic attractive interactions. We observe
the amplification of a local instability by seeding a momentum
state $\bf{p}$ and suddenly switching the scattering length
negative via a Feshbach resonance. As required by momentum
conservation, we also observe the appearance of atoms in the
conjugate momentum state. The time scale for this depletion
process is found to be comparable to that for global collapse,
implying that this process will be the primary decay channel for
large homogeneous condensates.
\end{abstract}

\pacs{03.75.Kk, 34.50.-s, 32.80.Pj}

\author{J. K. Chin}
\author{J. M. Vogels}
\author{W. Ketterle}

\homepage[Group website: ]{http://cua.mit.edu/ketterle_group/}
\affiliation{Department of Physics, MIT-Harvard Center for
Ultracold Atoms, and Research Laboratory of Electronics,
Massachusetts Institute of Technology, Cambridge, MA 02139}
\maketitle

\catcode`\ä = \active \catcode`\ö = \active \catcode`\ü = \active
\catcode`\Ä = \active \catcode`\Ö = \active \catcode`\Ü = \active
\catcode`\ß = \active \catcode`\é = \active \catcode`\è = \active
\catcode`\ë = \active \catcode`\ô = \active \catcode`\ê = \active
\catcode`\ø = \active \catcode`\ò = \active \catcode`\í = \active
\defä{\"a} \defö{\"o} \defü{\"u} \defÄ{\"A} \defÖ{\"O} \defÜ{\"U} \defß{\ss} \defé{\'{e}}
\defè{\`{e}} \defë{\"{e}} \defô{\^{o}} \defê{\^{e}} \defø{\o} \defò{\`{o}} \defí{\'{i}}%

Our current understanding of the collapse of Bose-Einstein
condensates (BEC's) with attractive interactions is incomplete.
While the experiments in $^7$Li provided many insights into the
formation kinetics and stability \cite{Gert00}, it was not until
the discovery of externally induced Feshbach resonances
\cite{Inou98} that it became possible to tune the value of the
scattering length and study in detail the effects of an attractive
mean field potential. This technique was used by Donley \textit{et
al} to study $^{85}$Rb with a negative scattering length, and they
have observed rich dynamics inherent in the collapse of the
condensate. Among the most intriguing observations was the
formation of low energy `bursts' and `jets' which were ejected out
of the condensate \cite{Donl01}. While theoretically the
enhancement of quantum fluctuations could give rise to such
phenomena \cite{Duin01, Yuro02, Calz02}, there is currently no
consensus on the exact mechanism by which it occurs.

So far, theoretical developments have been limited by having only
one experimental testing ground. The experiments using $^{85}$Rb
were done using small condensates ($\sim 15000$ atoms), where the
attractive mean field energy $\mu$ is comparable to or less than
the $\hbar\omega$ level spacing of the harmonic trapping
potential. In this paper, we study the collapse of large sodium
condensates far in the Thomas Fermi regime ($|\mu|
>> \hbar\omega$), where the spatial profile of the condensate is
relatively homogeneous. Much of the dynamics of such a system is
then described by local phenomena. When the interactions become
attractive, Yurovsky \cite{Yuro02} predicts that local
instabilities with momentum on the order of the (imaginary) speed
of sound will undergo exponential growth. Simultaneously, momentum
conservation requires atoms to be generated in conjugate momentum
states. Since amplification happens on the time scale of the
chemical potential $h/\mu$, the resulting quantum evaporation of
the zero momentum condensate atoms can happen faster than the
global collapse for large condensates, where the whole condensate
`implodes'.

We probe this decay channel by seeding a particular momentum state
with an initial population, then suddenly switching the
scattering length negative via a Feshbach resonance. At the same
time, the trapping potential is turned off so all subsequent
dynamics are due only to the intrinsic attractive interactions.
The resulting amplification and the associated generation of atoms
in the conjugate momentum state verifies the theory. We end with a
discussion on the different collapse time scales of competing
processes and show that for large condensates, this decay channel
becomes dominant.

The theoretical basis for the amplification of local instabilities
is the dispersion relation for the elementary excitations in a
Bose-Einstein condensate:
\begin{eqnarray}
\epsilon(\textbf{p})=\sqrt{\frac{|\textbf{p}|^2}{2m}(2n_0U+\frac{|\textbf{p}|^2}{2m})}
\label{eq:dispersion}
\end{eqnarray}
where $n_0$ is the density, $U=4\pi\hbar^2 a/m=\mu/n_0$ is the
contact potential, $a$ is the scattering length and $m$ is the
mass. For an elementary excitation whose momentum satisfies
$|\textbf{p}|^2/2m < 2|\mu|$, an instability forms when $\mu < 0$
(i.e. $a<0$) and oscillatory behavior gives way to exponential
growth or decay. A formal derivation gives the evolution of these
low momentum modes as
\begin{eqnarray}
\langle \xi^{\dag}_{\textbf{p}} \xi_{\textbf{p}} \rangle(t) &=&
\frac{|Un_0|^2}{\hbar^2\lambda^2(\textbf{p})}\sinh^2[\lambda(\textbf{p})t]
\label{eq:growth}
\end{eqnarray}
where $\lambda$(\textbf{p}) is given by
$|\epsilon$(\textbf{p})$|/\hbar$ (Eq \ref{eq:dispersion}) when
$a<0$ \cite{Yuro02} and $\xi_{\textbf{p}}$ is the destruction
operator for mode \textbf{p}. The instability of the mode pair
($\textbf{p}$, $-\textbf{p}$) results in correlated growth, where
the creation of an atom in the $+\textbf{p}$ state is accompanied
by the creation of an atom in the $-\textbf{p}$ state. A similar
phenomenon is also responsible for the four wave mixing process
observed in \cite{Deng99,Voge02}. At higher momentum, the energy
becomes real again as the excitations now have enough kinetic
energy to stabilize them against the attractive interactions.

In our experiments, we created large cigar-shaped sodium
condensates in the $F=1$, $m_{F}=-1$ spin state with typical atom
numbers of $\sim$30 million and peak densities of $3\times
10^{14}$ cm$^{-3}$ in a Ioffe-Pritchard (IP) magnetic trap.
Following this, they were adiabatically loaded into a mode-matched
1064 nm cylindrical optical dipole trap with trapping frequencies
of 250 Hz by 2 Hz and held for 1 s to allow transient excitations
to damp out. In the next 117 ms, the axial magnetic field was
ramped up to an intermediate value just short of the 1195 G
Feshbach resonance \cite{Inou98}. In order to obtain a homogeneous
field across the condensates, we used two pairs of coils coaxial
with the condensates. The first pair (bias coils) provided a large
bias field ($\sim$1200 G) with some gradient ($\sim$6 G/cm). The
second pair (pinch coils) produced a cancelling gradient together
with a small bias field ($\sim$8 G), resulting in a nominally
uniform bias field. The scaling of the field with current was
characterized at low fields, allowing us to adjust our fields to
within 0.4 G. At this point, the condensates were slightly
compressed by a small axial magnetic field curvature, giving them
radial and axial dimensions of $\sim15$ $\mu$m and 1.5 mm
respectively and a speed of sound $c=\sqrt{Un_0/m}$ of 8 mm/s.

We imprinted a low momentum excitation onto a stable condensate
($a>0$) using a two photon optical Bragg transition to couple
atoms from the zero momentum state to a low momentum state
$\bf{p}$ \cite{Sten99}. The Bragg beams were pulsed on for 400
$\mu$s and were directed at an angle of 15$^{\circ}$ and
20$^{\circ}$ respectively from the long axis of the condensate,
creating phonons propagating with a momentum of $m\times$2.5 mm/s
in a predominantly radial direction. With this momentum $\bf{p}$,
the minimum value of $|a|$ required for amplification to happen is
estimated to be $-0.06$ nm. The beams were red-detuned from the
sodium D$_2$ line by 3 nm to minimize Rayleigh scattering, and had
a frequency difference of $\Delta\omega=2 \pi \times 700$ Hz.

At this point, we changed the scattering length $a$ almost
instantaneously by switching off the small pinch coils, thereby
stepping up the magnetic field by 8 G and entering the Feshbach
resonance. This gave us a well-defined initial condition similar
to that in \cite{Donl01}. We chose to use the 1195 G Feshbach
resonance as it is the broadest resonance in sodium with a width
of $\sim 4$ G and the scattering length crosses zero at a magnetic
field below the singularity. This allowed us to tune the value of
$a$ continuously from the background scattering length $a_0$ to
zero to arbitrary negative values by varying the intermediate
field value at the end of the slow initial ramp. Since the pinch
coils are much smaller than the bias coils, their mutual
inductance is small and the sudden switch off only affected the
bias field adversely by $\leq 1$ G. Moreover, the sudden jump
reduced the time available for three-body decay, thereby
overcoming the problematic high loss rate first observed by Inouye
\textit{et al} \cite{Inou98}. We also turned off the optical
trapping potential in the same instant, allowing the system to
evolve under its intrinsic attractive interactions. After a
variable time of 0 to 1 ms, the magnetic fields were turned off,
causing the scattering length to become positive again. A pump
pulse from $F=1$ to $F'=2$ was then applied and radial absorption
images on the $F=2$ to $F'=3$ cycling transition were taken after
6 ms of ballistic expansion. These images provided the momentum
distribution of the condensate.

\begin{figure}[tbp]
\includegraphics[width=80mm]{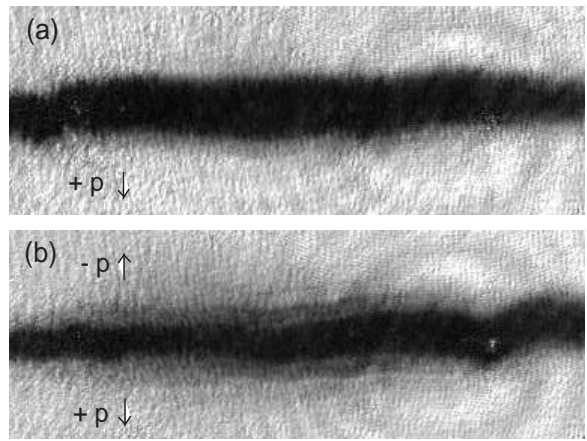}
\caption{Amplification of unstable excitations in a BEC with
$a<0$. In (a), the unamplified $+\bf{p}$ excitations are shown
moving out of the condensate after 6 ms ballistic expansion. In
(b), the $+\bf{p}$ excitations have been amplified and are
accompanied by the formation of $-\bf{p}$ excitations after 600
$\mu$s of hold time at $a=-0.82$ nm. The field of view is 960
$\mu$m by 100 $\mu$m.} \label{fig:phonons}
\end{figure}

The images in Fig.\ \ref{fig:phonons} probe the radial dynamics of
the condensates and provide dramatic visual verification of
quantum evaporation. Fig.\ \ref{fig:phonons}(a) shows the
$+\bf{p}$ excitations moving out of the condensate without any
amplification. In contrast, Fig.\ \ref{fig:phonons}(b) was taken
after the condensate had been held at $a=-0.82$ nm for 600 $\mu$s.
Not only was the number of atoms in the $+\bf{p}$ momentum state
significantly amplified, it was accompanied by the formation of
excitations in the $-\bf{p}$ momentum state, here seen moving out
of the condensate in the opposite direction. These observations
clearly demonstrate the instability of a condensate with negative
scattering length.

Due to the large aspect ratio of our condensates, only part of the
condensate could be imaged at high magnification. However, since
the $+\bf{p}$ and $-\bf{p}$ excitations were created predominantly
in the radial direction, this was not a limitation. The small
`kinks' that were also apparent in our condensate are most likely
a result of imperfections in our optical dipole trap. Yet rather
than degrade the signal, they highlight the parallel contours
between the condensate and the ridge of excitations as atoms move
out with a definite momentum.

In order to perform more quantitative tests of this phenomenon, we
first characterized the negative scattering length dependence on
the field by directly probing the strength of the attractive
interactions. To do this, we prepared optically trapped
condensates close to a Feshbach resonance as above. The confining
infrared laser beam was then replaced with a repulsive 3 nm
blue-detuned ``antitrap'' beam as we simultaneously jumped to
negative scattering lengths. At the right intensity, the antitrap
beam provided the correct amount of repulsive dipole force needed
to compensate for the attractive interactions within the
condensate and suppressed any global contraction of size. However,
this is an unstable equilibrium and any sloshing of the condensate
or misalignment of the laser beam caused the condensate to be
repelled. Therefore the antitrap was fine-aligned to milliradian
accuracy such that a condensate with $a>0$ was ripped apart
radially into a hollow cylinder. Using this method, we were able
to stabilize an attractive condensate significantly for 0.2 to 2
ms, depending on $a(B)$, before unavoidable losses became
significant. The radial dimension was used to monitor the
mechanical dynamics of the condensate occurring on the 250 Hz time
scale of the trap frequency. For minimal distortion of the spatial
image, absorption images were taken after only 2 ms of ballistic
expansion necessary for the high magnetic fields to die out.

\begin{figure}[tbp]
\includegraphics[width=60mm]{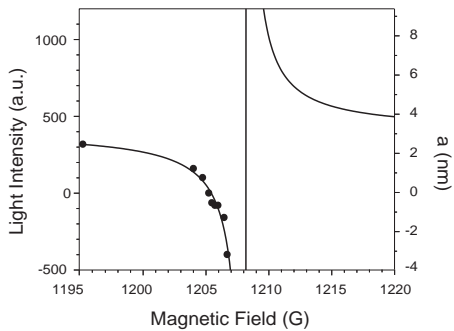}
\caption{Characterization of the Feshbach resonance. The light
intensity needed to balance out the interatomic interactions gives
the dependence of the scattering length on magnetic field. The
width of the resonance was found to be
2.4$\pm0.4$ G.} \label{fig:feshbach}
\end{figure}

In equilibrium, $a\varpropto
F_{\mathrm{attractive}}=F_{\mathrm{repulsive}}\varpropto I$.
Therefore we plot the light intensity $I$ needed for stabilization
vs.\ magnetic field $B$ in Fig.\ \ref{fig:feshbach} and obtain the
scattering length dependence on the field. A red-detuned laser
beam with a similar detuning of 3 nm was employed to obtain the
points in the positive scattering length regime. By fitting the
expected Feshbach curve $a(B)=a_0(1+\frac{\Delta B}{B-B_{0}})$ to
our data, we find the width $\Delta B$ of the 1195 G Feshbach
resonance to be (2.4$\pm 0.4$) G. Here, $a_0=3.3$ nm is the
triplet scattering length at high fields \cite{Samu00}. While the
exact position $B_{0}$ of the resonance is unimportant for our
experiment, the width of the resonance is crucial as it determines
the range of the magnetic field we have to work within
\cite{foot}.

\begin{figure}[tbp]
\includegraphics[width=60mm]{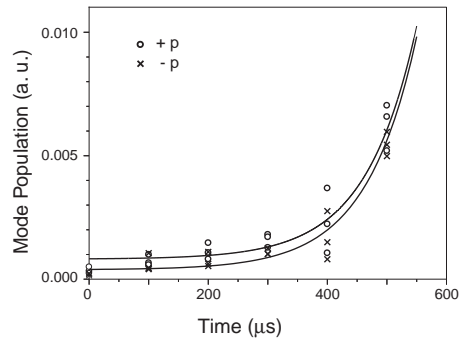}
\caption{Amplification of excitations in a BEC with $a<0$. The
growth of the normalized mode population in both $+\textbf{p}$ and
$-\textbf{p}$ modes as a function of hold time at $a=-1.35$ nm is
fitted to an exponential dependence according to Eq.\
\ref{eq:growth}, which gave a common growth rate of 5.89$\pm$0.83
ms$^{-1}$. A variable offset was allowed to account for the
initial seeding of the $+\textbf{p}$ mode.} \label{fig:growth}
\end{figure}

A quantitative analysis of the growth in the $+\bf{p}$ and
$-\bf{p}$ modes was performed by monitoring their occupation
number as a function of hold time in the attractive regime (Fig.\
\ref{fig:growth}). Taking into account the high loss of atoms
during this process due to three-body decay or inelastic two-body
collisions, we normalize the number count for each mode to the
number of atoms in the condensate at the end of the hold time. The
maximum duration of amplification was limited by the lifetime of
the condensate, which was about 600 $\mu$s for $a=-1.35$ nm.
Following Eq.\ \ref{eq:growth}, an exponential dependence was
fitted to the data, yielding a common growth rate of 5.89$\pm$0.83
ms$^{-1}$ for both modes. This agrees well with the theoretical
growth rate of 5.57 ms$^{-1}$, estimated using our initial mean
field of $h\times5$ kHz.

\begin{figure}[tbp]
\includegraphics[width=60mm]{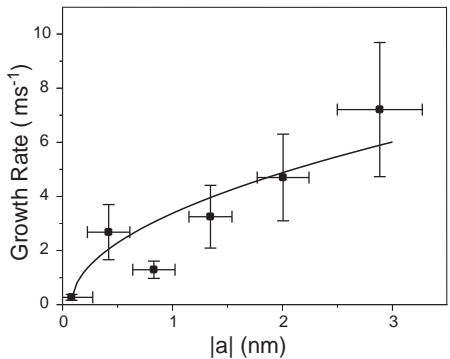}
\caption{Growth rate of excitations in a BEC with $a<0$. The rate
increases with the magnitude of the negative scattering length
$|a|$. The solid line shows a best fit to the predicted
$b_1\sqrt{|a|-b_2}$ dependence.} \label{fig:a_dependence}
\end{figure}

We also investigated the dependence of the growth rate on the
magnitude of the negative scattering length $|a|$. By varying the
scattering length and extracting the growth rate of the
excitations as above, we observed a strong increase of the growth
rate with $|a|$ (Fig.\ \ref{fig:a_dependence}). A fit to the
theoretical prediction $\lambda(\textbf{p}) = b_1\sqrt{|a|-b_2}$
(Eq.\ \ref{eq:dispersion}) yielded the fit parameters
$b_1=(1.11\pm0.14)\times 10^8$ nm$^{1/2}$ms$^{-1}$ and
$b_2=0.078\pm0.04$ nm, in agreement with theoretical estimates of
1.55 $\times 10^8$ nm$^{1/2}$ms$^{-1}$ and 0.06 nm respectively.
For this sequence of measurements, the initial mean field was
$\mu=h\times4$ kHz. The large error bars reflect the high
sensitivity of the dynamics to the magnetic field. In particular,
the three-body recombination and inelastic loss rates as a
function of $a$ have not yet been well characterized, which limits
the accuracy of our data.

The results presented here prove conclusively that quantum
evaporation is a part of the complex dynamics that take place
during the collapse of an attractive condensate. While we select a
particular mode for observation, the effect is predicted to happen
for all modes satisfying the condition $|\textbf{p}|^2/2m <
2|\mu|$. At short times, the pairwise emission of atoms also
implies that the number of atoms in the conjugate mode pairs will
be exactly correlated, although as the condensate becomes
increasingly depleted, higher order effects will degrade the
correlation \cite{Yuro02}.

\begin{figure}[tbp]
\includegraphics[width=60mm]{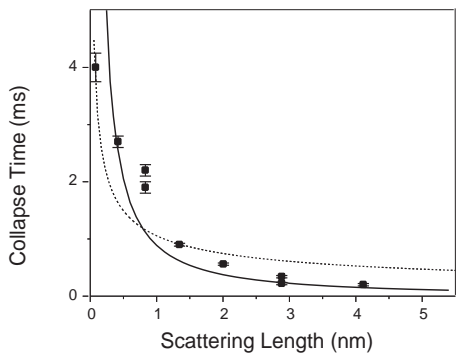}
\caption{Time scales for BEC decay. The observed collapse time is
compared to the time taken for the condensate density to fall to
1/$e$ of its original value ($t_{1/e}$) due to instability
amplification (solid curve, Eq.\ \ref{eq:t_1/e}) and the expected
$t_{\mathrm{decay}}$ for global collapse (dashed curve, Eq.\
\ref{eq:t_decay}).} \label{fig:decay}
\end{figure}

Since quantum evaporation is intrinsic to the condensate, a
natural question to ask would be how large a part it plays in the
unperturbed collapse of the condensate. We study this by comparing
the observed time taken for the condensate to decay completely
without initial seeding (Fig.\ \ref{fig:decay}), with the
theoretical prediction \cite{Yuro02} for the depletion of $n_0$:
\begin{eqnarray}
\dot{n_0}=-4\sqrt{\frac{2\pi\hbar}{mt}}a^2n^2_0\exp(8\pi\hbar|a|n_0t/m)
\label{eq:t_1/e}
\end{eqnarray}
From Eq.\ \ref{eq:t_1/e}, we extract the time $t_{1/e}$ taken for
the condensate density $n_0$ to decay to $1/e$ of its original
value with our parameters (Fig.\ \ref{fig:decay}, solid line). As
both the number of unstable modes and the amplification rate
increase with $a$, there is a significant decrease of $t_{1/e}$
with $|a|$. We also considered the effect of the initial quantum
depletion ($\sim 1\%$, \cite{Huang}) on $t_{1/e}$, but concluded
that its impact is small.

In addition, we compare the observed collapse time to the decay
time predicted for global collapse. Even for condensates initially
far in the Thomas-Fermi regime, their spatial profile is not
completely uniform and the inherent pressure gradient will cause
the condensate to collapse inwards. As the condensate compresses,
the density will increase and sharply enhance the three-body
recombination loss rate, which goes like $n_0^2$. We model the
radial evolution using the root mean square radius $R=\int
r^2|\psi(r)|^2 rdr$. For our cylindrical condensates with an
aspect ratio of 100:1, an analytical solution for the resulting 2D
dynamics exists \cite{Yu96, Yu97}, given by
$\ddot{R}=4(\frac{E}{m}-\omega^2 R)$ where $E$ is the total energy
of the system. Since $\omega=0$ and $a$ is constant throughout in
our experiments, $E$ is conserved and obtained from initial
conditions. The time taken to reach $R=0$ is
\begin{eqnarray}
t_{\mathrm{decay}}=\frac{1}{\omega_0}\sqrt{\frac{a_0}{|a|}}
\label{eq:t_decay}
\end{eqnarray}

Eqs.\ \ref{eq:t_1/e} and \ref{eq:t_decay} are plotted in Fig.\
\ref{fig:decay} and their intersection separates the graph into
two domains. In the first, $a$ is small and global collapse is
predicted to dominate over quantum evaporation. For higher $a$'s,
the converse is true as amplification rate and the number of
unstable modes increase. The comparison with the data suggests
that the decay of the condensate is not dominated by three-body
decay, but is caused by amplification of unstable modes. We were
unable to study the condensate lifetime at even more negative
scattering lengths since the condensate then decayed almost
instantaneously. In order to further separate the two time scales,
the global collapse time needs to be much slower compared to
$t_{1/e}$. This would require a combination of a weak trap with a
high number of atoms, which is currently out of reach.

In conclusion, we have shown that large condensates far in the
Thomas Fermi regime undergo amplification of local instabilities
when their scattering length becomes negative. We have studied the
dependence of amplification rate on the magnitude of
the negative scattering length and found reasonable agreement with
the theory. For our parameters, this quantum evaporation process
becomes comparable or faster than the global collapse.

We thank K.\ Xu, T.\ Mukaiyama and J.\ R.\ Abo-Shaeer for
experimental assistance. We would also like to acknowledge J.\
Anglin for valuable comments, as well as A.\ E.\ Leanhardt, and
Z.\ Hadzibabic for critical reading. This work was funded by ONR,
NSF, ARO, NASA, and the David and Lucile Packard Foundation.

\bibliographystyle{apsrev}

\begin{thebibliography}{14}
\bibitem[{foo()}]{foot}
\bibinfo{note}{We observe the resonance around 1208$\pm$10 G, which would agree
  with the theoretical value within the systematic error.}
\end{thebibliography}

\end{document}